# High Cadence Millimagnitude Photometric Observation of V1112 Persei (Nova Per 2020)

Neil Thomas, Kyle Ziegler and Peter Liu
Department of Astronautical Engineering, United States Air Force Academy, CO 80840, USA;
neil.thomas@afacademy.af.edu

**Abstract**

The private Lookout Observatory (LO) monitored the classic nova V1112 Persei on 37 nights spanning over 80 days, beginning shortly after its discovery by Seiji Ueda on 25 Nov 2020. Images were captured at a high cadence, with exposure lengths of initially less than 2 seconds and with some sessions lasting more than ten hours. The standard error of the photometry was typically better than 5 thousandths of a magnitude (5 mmag). This cadence and precision allowed for not only the observation of the expected dimming of the nova, but also variability having a period of 0.608 +- 0.005 days. This data compliments the publicly available photometry from the American Association of Variable Star Observers (AAVSO) and the resultant data is combined to perform this photometric analysis. This paper does not attempt an in-depth physical analysis of the nova from an astrophysical perspective.

1. Introduction

Classic novae are close proximity binary stars comprised of a white dwarf primary and a (typically) late-type main sequence star (Warner 1995). The large mass of the white dwarf combined with their close pairing results in a deformed secondary that consistently loses material to the primary (Bode and Evans 2008). This material either forms an accretion disk or simply flows directly to the surface of the primary, depending on whether the primary contains a magnetic field. Eventually a critical mass of material accumulates on the surface of the white dwarf resulting in a runaway hydrogen fusion reaction, or nova.

V1112 Persei was discovered on 25 November 2020 (Ueda 2020) at an unfiltered magnitude of 10.6. Spectroscopy obtained by Munari et al. on 26 November identified the nova as a galactic nova near maximum at the time of first observation.

A campaign was developed to obtain photometry on this nova starting three days after its discovery. The goals were to both record the long-term evolution of its brightness and to leverage the high cadence and mmag precision of the LO configuration to search for short-term variations.

Section 2 details the observations. Section 3 discusses the LO instrumentation and data reduction methods. Section 4 describes the photometric analysis of this nova, while Section 5 is the conclusion.

## 2. Observing Campaign

V1112 Persei was observed over 37 nights between 25 Nov 2020 and 16 Feb 2021 and a total of nearly 190 hours of target photometry were collected. Exposures ranged from 2 seconds in November up to 90 seconds by the end of the survey. The dates, duration, and mean nightly magnitudes of the observations are listed in Table 1. Individual LO measurements can be obtained from the AAVSO website (Observer Code TNBA).

Table 1. Observation Details

| Date | Duration (hours) | Mag (Gaia G) | Mag err | Date | Duration (hours) | Mag (Gaia G) | Mag err |
|---|---|---|---|---|---|---|---|
| 28 Nov 2020 | 5.80 | 8.252 | 0.0119 | 25 Dec 2020 | 6.23 | 9.778 | 0.0219 |
| 29 Nov 2020 | 10.44 | 8.689 | 0.0186 | 26 Dec 2020 | 10.19 | 9.747 | 0.0090 |
| 30 Nov 2020 | 2.77 | 8.341 | 0.0202 | 27 Dec 2020 | 3.02 | 9.854 | 0.0120 |
| 1 Dec 2020 | 2.98 | 8.271 | 0.0196 | 30 Dec 2020 | 8.90 | 9.800 | 0.0115 |
| 4 Dec 2020 | 1.42 | 8.125 | 0.0109 | 31 Dec 2020 | 7.80 | 10.056 | 0.0173 |
| 5 Dec 2020 | 10.17 | 8.639 | 0.0218 | 2 Jan 2021 | 4.59 | 10.242 | 0.0233 |
| 6 Dec 2020 | 9.60 | 8.525 | 0.0257 | 5 Jan 2021 | 3.03 | 10.638 | 0.0151 |
| 7 Dec 2020 | 10.18 | 8.294 | 0.0275 | 7 Jan 2021 | 2.23 | 10.024 | 0.0306 |
| 8 Dec 2020 | 10.18 | 8.503 | 0.0314 | 11 Jan 2021 | 3.80 | 12.424 | 0.0176 |
| 9 Dec 2020 | 9.40 | 8.874 | 0.0159 | 15 Jan 2021 | 5.06 | 13.083 | 0.0178 |
| 10 Dec 2020 | 8.50 | 9.020 | 0.0178 | 20 Jan 2021 | 0.90 | 13.612 | 0.0530 |
| 14 Dec 2020 | 2.12 | 8.878 | 0.0598 | 21 Jan 2021 | 0.64 | 13.380 | 0.0255 |
| 16 Dec 2020 | 9.90 | 9.049 | 0.0324 | 23 Jan 2021 | 2.10 | 13.865 | 0.0580 |
| 17 Dec 2020 | 9.80 | 9.268 | 0.0172 | 24 Jan 2021 | 2.57 | 13.955 | 0.0997 |
| 20 Dec 2020 | 1.24 | 9.244 | 0.0103 | 28 Jan 2021 | 3.32 | 14.243 | 0.0630 |
| 22 Dec 2020 | 9.59 | 9.573 | 0.0141 | 29 Jan 2021 | 4.03 | 14.367 | 0.0660 |
| 23 Dec 2020 | 1.24 | 9.601 | 0.0070 | 30 Jan 2021 | 0.52 | 14.471 | 0.0830 |
| 24 Dec 2020 | 1.00 | 9.686 | 0.0146 | 8 Feb 2021 | 3.16 | 14.951 | 0.0507 |
| | | | | 16 Feb 2021 | 2.30 | 15.133 | 0.0587 |

## 3. Instrumentation and Methods

The LO is primarily optimized to maintain the photometric precision necessary to observe exoplanet transits. It consists of an 11" Celestron telescope modified to f/1.9 with a HyperStar. A ZWO ASI 1600 CMOS camera performs the imaging, and optical filters are not typically used. Dawn and dusk flats are captured and applied in the normal way. During favorable conditions, individual measurements typically have noise in the 10-20 mmag range. This is primarily white noise, however, and differential photometry with noise levels of less than 2 mmag is often maintained from dusk to dawn by binning data to simulate 2–4-minute exposures. Absolute magnitude calibration between nights is generally consistent to within 20 mmag.

Images are collected semi-autonomously using MAXIM DL and CCDCOMMANDER. Additionally, a custom-made software pipeline developed in MATLAB performs the aperture photometry.

Although a full discussion of the software pipeline's design is not currently available, Thomas and Guan (2020) provide a somewhat more details performance analysis.

Calibration stars are automatically selected based on similarity of magnitude and color. This field of view (FOV) provided approximately 1,000 field stars at the beginning of the campaign and 9,000 at the end, as exposure lengths increased. Typically, the most compatible 50-150 stars are automatically selected and used for differential photometry. This provides relative magnitudes, not the absolute magnitudes required to study this nova over several months. To derive absolute magnitudes, we use Gaia DR2 magnitudes and colors to derive absolute Gaia G-band magnitudes for all usable field stars (Gaia Colab. et al. 2016, 2018). To do this, the instrumental magnitude of each star is compared to its Gaia magnitude and used to determine the first order shift to true magnitudes. Then a second order color correction is applied after fitting the magnitude residuals to Gaia B-V colors. Although our photometry is unfiltered and Gaia is G-band, a reliable transformation is possible. Our calibrated magnitudes are compared to Gaia values for a typical night in Fig. 1. Although noise is photon dominated for faint stars, the standard deviation of the difference between our values and Gaia among the brightest third of stars is 15 mmag. For comparison, the quoted errors of Gaia magnitudes in this brightness range are typically 2-6 mmag. It may seem odd to convert our clear unfiltered (CV) data into the Gaia G-band and then to eventually compare those results to AAVSO CV and visual band (V) photometry. We do this because of our overall desire to automatically select and use many calibration stars from among the entire FOV. Usable field stars extend down to $17^{th}$ magnitude and the authors know of no catalog besides Gaia containing magnitude and color information to this faintness. Additionally, attempting to remain in our native CV would still present compatibility complications since unfiltered data is not a true band because observed flux will depend on sensor sensitivity at various wavelength. This calibration method will, however, introduce a source of error during the second order color correction for this target. We use the B-R value of 0.80776 obtained by Gaia prior to the nova's outburst. But in reality, we expect a reddening of the nova during its evolution (Woudt and Ribeiro 2014). Since this color evolution is not known beforehand, we use this fixed value. We will show later that these color correction errors are not overwhelming, especially for our primary objective of analyzing the short-period signal.

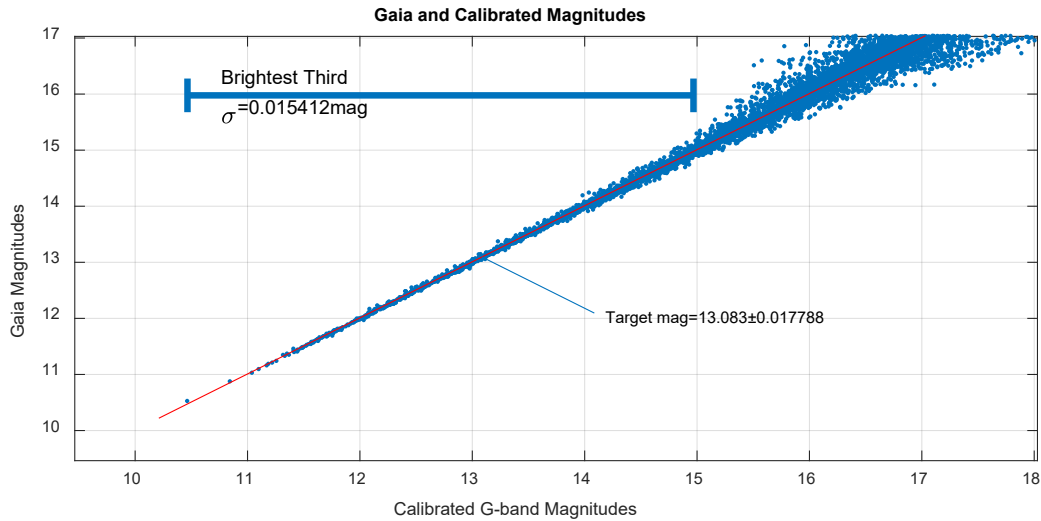

Figure 1: Performance of calibration from unfiltered (CV) to Gaia G-band on a typical night (15 Jan 2021). Approximately 9,000 field stars were transformed to absolute magnitudes. Precision deteriorates for faint stars but appears to be driven by shot noise with no obvious systematic influences. The standard deviation between computed magnitudes and Gaia values is 0.015 mag for the brightest third of stars. This is only about 0.010 mag greater than the errors in the Gaia values themselves and is acceptable for calibrating our data across different nights.

**4. Analysis and Results**

AAVSO data was retrieved from the AAVSO International Database on 23 January 2021 and included 44,608 measurements. AAVSO photometry comes from many sources and through a variety of filters. Most (~80%) observations of this target are in CV and in Johnson V (V). The data from the LO, although initially collected as unfiltered, is calibrated to Gaia green (G) and provides an additional 65,462 measurements.

A simple transformation between photometry taken in different colors can usually be determined for a given star. But since this nova is likely changing color over time, making different bands of photometry compatible must be approached more carefully. The light curves in AAVSO's CV and our G-calibrated are shown in Fig. 2. The CV and G bands appear highly compatible. There are very few periods in which there are overlapping observations with which to generate a transform. During those few overlapping periods of data we see a disagreement on the order of 0.03 mag. Furthermore, we see similar levels of disagreement between simultaneous AAVSO CV measurements reported by different observers. We therefore choose to incorporate AAVSO CV photometry as directly equivalent to our G-band data.

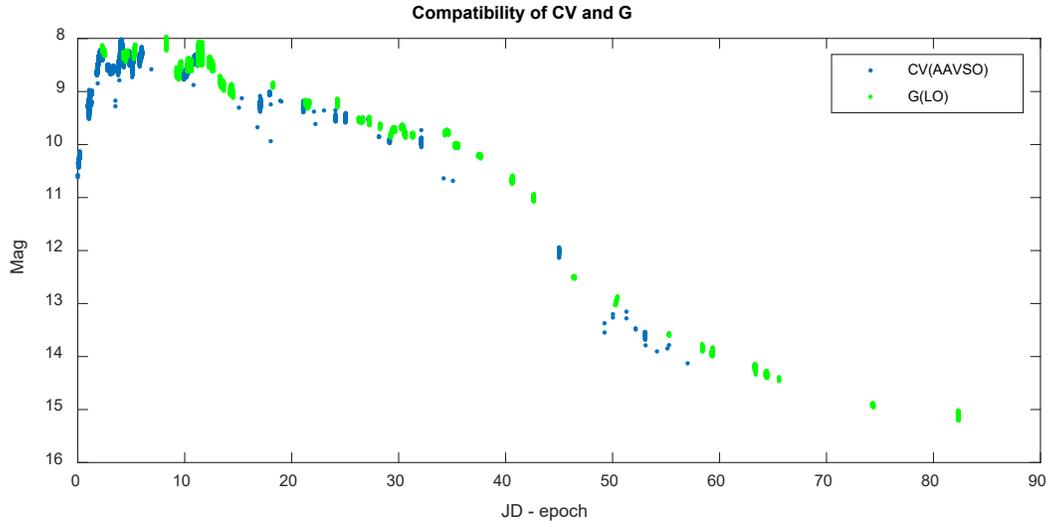

Figure 2: Photometric observations of V1112 in CV (from AAVSO) and G (our results). CV and G bands are compatible with each other with no transformation being needed. The epoch is the time of the first observation available from AAVSO (2459179.34333)

The majority of AAVSO data is in the V-band and our data (and the relatively equivalent AAVSO CV data) requires a transformation. In Fig. 3 we see the divergence with time between G/CV and V bands. The V-band brightness fades faster than the G/CV and a time dependent transformation is required to combine the data.

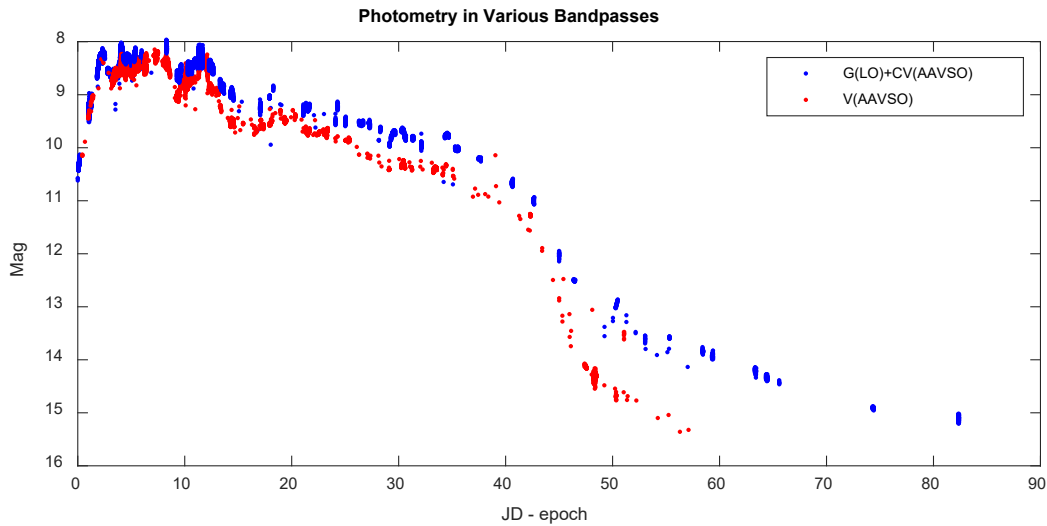

Figure 3: The V-band observations fade more drastically with time than of the G/CV measurements due to the changing color of the nova, requiring a transformation before combining the data.

To make these sources compatible, photometry in each color is binned into equal intervals by creating two-minute time steps over the entire time range. All data of a particular color within each interval is averaged to create a single photometric measurement at that color. The deviations between different colors during each interval can then be directly compared since they now have the same time sampling. The deviation between these magnitudes is shown in Fig. 4.

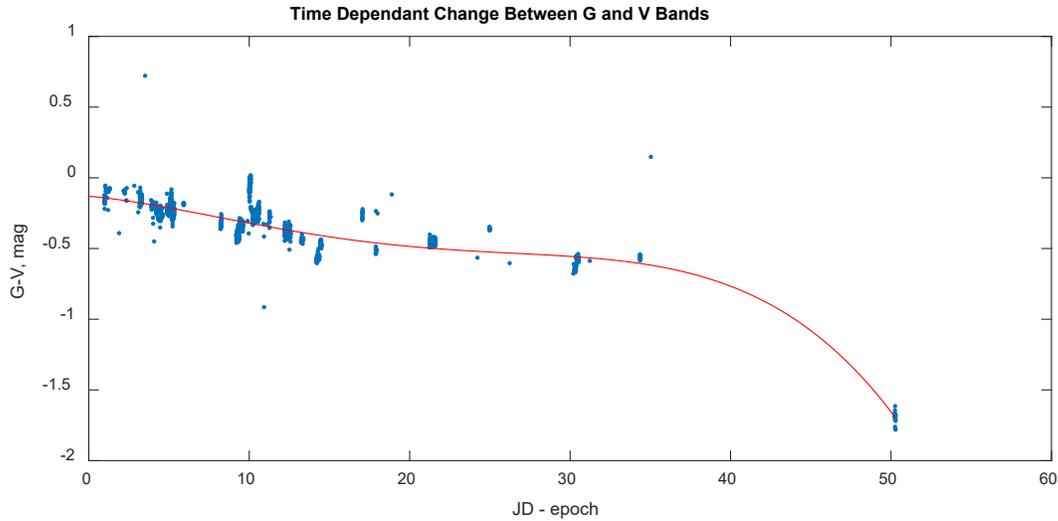

Figure 4: The divergence between G and V-band photometry over time due to the changing color of the nova. A fourth order polynomial fit is also shown and used to transform G-band data into V-band.

The transformation is given as a function time by the following equation, where JD' is the number of days past epoch, G is the observed magnitude in G-band, and V is the transformed magnitude to the V-band.

$$V = G - (-1.262x10^{-6}JD'^4 + 8.716x10^{-5}JD'^3 - 1.604x10^{-3}JD'^2 - 1.042x10^{-2}xJD' - 0.131)$$

The transformed light curve from all sources is shown in Fig. 5 for the first ten days after epoch. All sources are now highly compatible and show a smooth continuum of photometry.

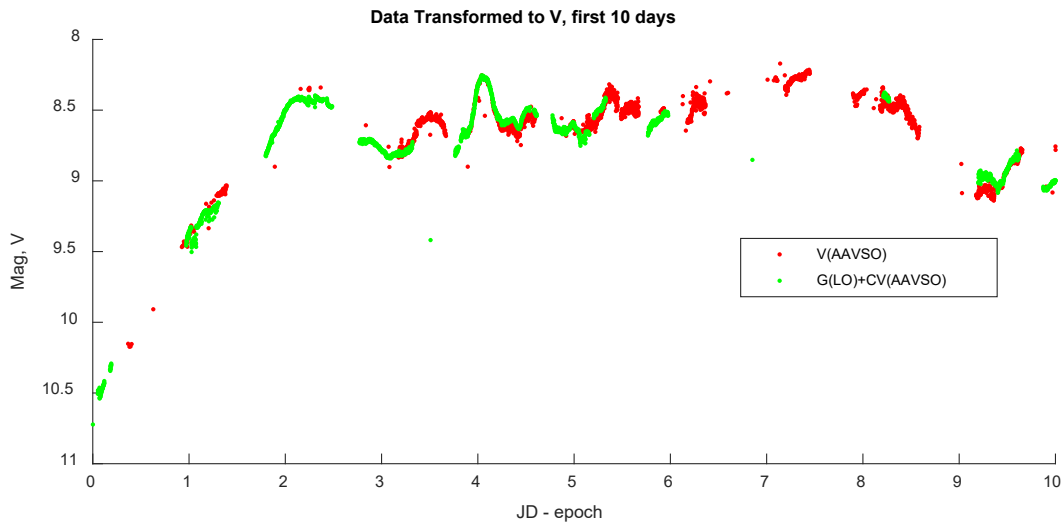

Figure 5: The first ten days of photometry after CV and G data are transformed to V using a time-dependent transform. The three sources are now highly compatible.

It is apparent in Fig. 5 that short-term variations well beyond the noise levels of the data are present. To determine the period of this activity, we first determine the long-term profile of the fading nova by finding its moving average with a smoothing interval of one day. The photometry and the smoothed fit are shown in Fig. 6.

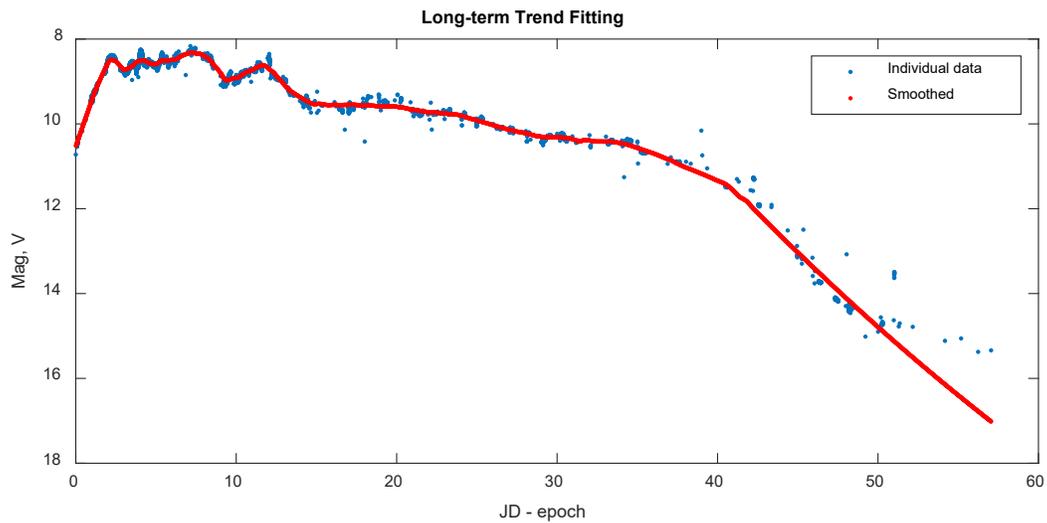

Figure 6: The long-term dimming of the nova (solid) is modelled by smoothing the data with intervals of one day. The short-term variation is then extracted by subtracting this long-term profile.

The long-term profile is subtracted from the photometry to isolate the variability as shown in Fig. 7. The periodic nature of the variability is apparent. Its amplitude ranges approximately a few hundred mmag but varies drastically, even between adjacent cycles.

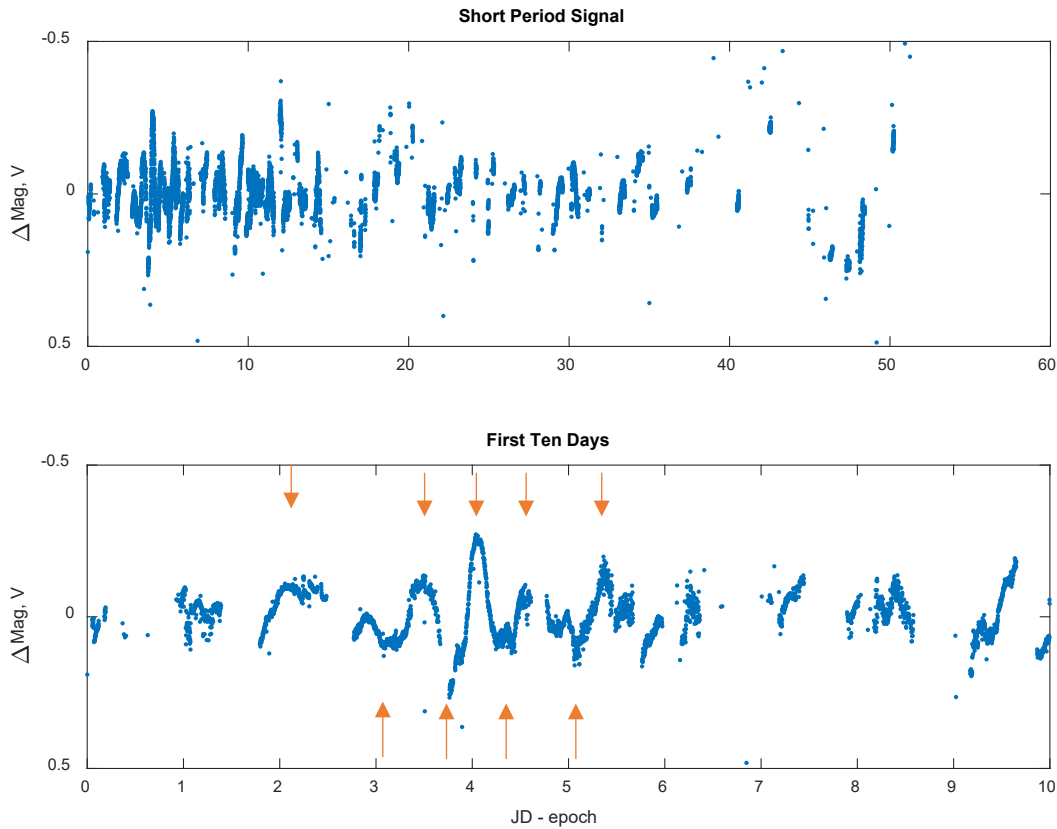

Figure 7: (Top) All photometry after being subtracted from the long-term brightness profile of the nova. (Bottom) Short-term variation in the first ten days. The periodic nature of this variation is clear, although its amplitude is inconsistent. Arrows point to peaks and troughs to highlight the periodic nature of the signal.

The Lomb-Scargle periodogram of this signal is given in Fig. 8, showing a convincing power peak at a period of 0.608 days (Lomb 1976). The Lomb-Scargle periodogram is a commonly used algorithm to detect periodic signals in unevenly spaced data. It is applied in MATLAB using the PLOMB.m function. It is based on Fourier transform theory and directly returns vectors of matching frequency and power values. A relatively high power at a given frequency indicates a repeating signal. We do not detect the much shorter period (0.09271 days) and smaller amplitude variability previously reported in the I-band (Schmidt 2021). This may not be surprising considering the difference in filters. We have some confidence in our results, however, as one can manually retrieve the 0.608-day period by measuring the peak-to-peak intervals shown in Fig. 7. To crudely estimate the error of our derived period we repeat this period search for only the first 20 days and then for days 20-40. The results are 0.6107 and 0.6060 days, respectively. If the period is not actually changing then our error is on the order of the difference between these independent results, or 0.005 days.

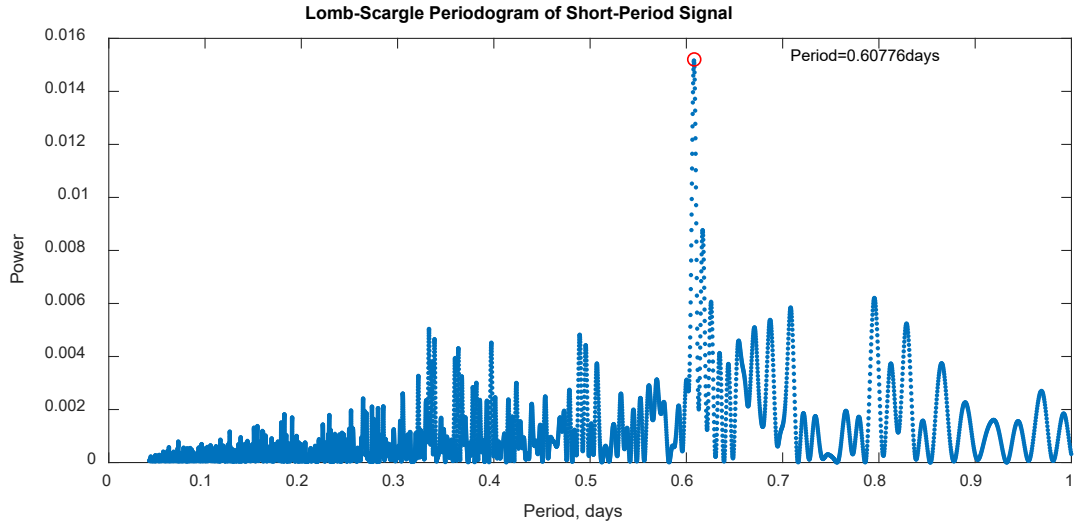

Figure 8: The Lomb-Scargle periodogram of short period magnitude variations. There is a clear peak at 0.608 days.

This is not a signal that responds well to phase folding due to its irregular amplitudes, as seen in Fig. 9. When the phased data is further binned using 200 observations per data point, however, a convincing sinusoidal form is presented with an amplitude of approximately 75 mmag.

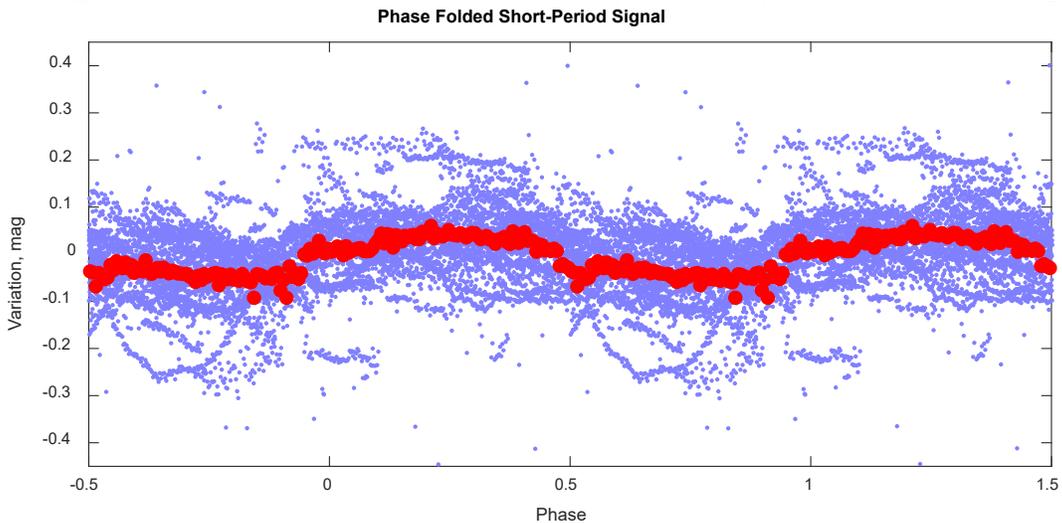

Figure 9: The short period signal phase folded to the detected period. Individual data (blue) is not particularly convincing. When binned by a factor of 200 (red), however, a convincing sinusoidal signal with a mean amplitude of approximately 75 mmag is apparent.

## 5. Conclusion

An 80-day campaign monitored the dimming of Nova Per 2020 and was combined with the AAVSO database to detect a visual variability having a period of 0.608 days and an irregular amplitude that ranges between approximately 50 and 200 mmag. These results may allow for a more detailed understanding of the physical processes at play in this nova.

**Acknowledgements**

We acknowledge with thanks the variable star observations from the *AAVSO International Database* contributed by observers worldwide and used in this research.

PA#: USAFA-DF-2021-104